\input harvmac
\overfullrule=0pt
\parindent 25pt
\tolerance=10000

%\draftmode
\lref\gibbonsb{G.~W.~Gibbons,
``Quantized Flux Tubes In Einstein-Maxwell Theory And Noncompact
Internal Spaces'',
Print-86-0411 (CAMBRIDGE)
{\it Presented at 22nd Karpacz Winter School of Theoretical Physics: Fields
  and Geometry, Karpacz, Poland, Feb 17 - Mar 1, 1986}.}

\lref\GarfinkleEQ{
D.~Garfinkle and A.~Strominger,
``Semiclassical Wheeler Wormhole Production,''
Phys.\ Lett.\ B {\bf 256}, 146 (1991).
%%CITATION = PHLTA,B256,146;%%
}

\lref\melvin{M.~A.~Melvin, ``Pure Magnetic and Electric Geons'',
Phys. Lett. {\bf 8} (1964) 65.}

\lref\ernst{F.~J.~Ernst, ``Removal of the nodal singularity of the C-metric'',
J.Math.Phys. {\bf 17} (1976) 515.}

\lref\gibbonsa{
G.~W.~Gibbons and K.~Maeda,
``Black Holes And Membranes In Higher Dimensional Theories With Dilaton
Fields,'' 
Nucl.\ Phys.\ B {\bf 298}, 741 (1988).
%%CITATION = NUPHA,B298,741;%%
}

\lref\WittenGJ{
E.~Witten,
``Instability Of The Kaluza-Klein Vacuum,''
Nucl.\ Phys.\ B {\bf 195}, 481 (1982).
%%CITATION = NUPHA,B195,481;%%
}

%\SaffinKY
\lref\SaffinKY{
P.~M.~Saffin,
``Gravitating fluxbranes,''
Phys.\ Rev.\ D {\bf 64}, 024014 (2001)
[arXiv:gr-qc/0104014].
%%CITATION = GR-QC 0104014;%%
}

\lref\GutperleMB{
M.~Gutperle and A.~Strominger,
 ``Fluxbranes in string theory,''
JHEP {\bf 0106}, 035 (2001)
[arXiv:hep-th/0104136].
%%CITATION = HEP-TH 0104136;%%
}

\lref\RussoTF{
J.~G.~Russo and A.~A.~Tseytlin,
 ``Magnetic backgrounds and tachyonic instabilities in closed superstring  
theory and M-theory,''
Nucl.\ Phys.\ B {\bf 611}, 93 (2001)
[arXiv:hep-th/0104238].
%%CITATION = HEP-TH 0104238;%%
}
\lref\CostaIF{
M.~S.~Costa, C.~A.~Herdeiro and L.~Cornalba,
 ``Flux-branes and the dielectric effect in string theory,''
arXiv:hep-th/0105023.
%%CITATION = HEP-TH 0105023;%%
}

\lref\EmparanRP{
R.~Emparan,
 ``Tubular branes in fluxbranes,''
Nucl.\ Phys.\ B {\bf 610}, 169 (2001)
[arXiv:hep-th/0105062].
%%CITATION = HEP-TH 0105062;%%
}

\lref\SparksGC{
J.~F.~Sparks,
 ``Kaluza-Klein branes,''
arXiv:hep-th/0105209.
%%CITATION = HEP-TH 0105209;%%
}

\lref\SaffinJG{
P.~M.~Saffin,
 ``Fluxbranes from p-branes,''
Phys.\ Rev.\ D {\bf 64}, 104008 (2001)
[arXiv:hep-th/0105220].
%%CITATION = HEP-TH 0105220;%%
}

\lref\BrecherXJ{
D.~Brecher and P.~M.~Saffin,
``A note on the supergravity description of dielectric branes,''
Nucl.\ Phys.\ B {\bf 613}, 218 (2001)
[arXiv:hep-th/0106206].
%%CITATION = HEP-TH 0106206;%%
}

\lref\MotlDJ{
L.~Motl,
``Melvin matrix models,''
arXiv:hep-th/0107002.
%%CITATION = HEP-TH 0107002;%%
}

\lref\SuyamaNE{
T.~Suyama,
``Melvin background in heterotic theories,''
arXiv:hep-th/0107116.
%%CITATION = HEP-TH 0107116;%%
}

\lref\UrangaDX{
A.~M.~Uranga,
``Wrapped fluxbranes,''
arXiv:hep-th/0108196.
%%CITATION = HEP-TH 0108196;%%
}

\lref\SuyamaGD{
T.~Suyama,
``Properties of string theory on Kaluza-Klein Melvin background,''
arXiv:hep-th/0110077.
%%CITATION = HEP-TH 0110077;%%
}

\lref\TakayanagiJJ{
T.~Takayanagi and T.~Uesugi,
``Orbifolds as Melvin geometry,''
arXiv:hep-th/0110099.
%%CITATION = HEP-TH 0110099;%%
}

\lref\RussoNA{
J.~G.~Russo and A.~A.~Tseytlin,
``Supersymmetric fluxbrane intersections and closed string tachyons,''
arXiv:hep-th/0110107.
%%CITATION = HEP-TH 0110107;%%
}

\lref\ChenNR{
C.~M.~Chen, D.~V.~Gal'tsov and P.~M.~Saffin,
``Supergravity fluxbranes in various dimensions,''
arXiv:hep-th/0110164.
%%CITATION = HEP-TH 0110164;%%
}

\lref\FigueroaNX{
J.~Figueroa-O'Farrill and J.~Simon,
``Generalized supersymmetric fluxbranes,''
arXiv:hep-th/0110170.
%%CITATION = HEP-TH 0110170;%%
}

\lref\DudasUX{
E.~Dudas and J.~Mourad,
``D-branes in string theory Melvin backgrounds,''
arXiv:hep-th/0110186.
%%CITATION = HEP-TH 0110186;%%
}

\lref\TakayanagiAJ{
T.~Takayanagi and T.~Uesugi,
``D-branes in Melvin background,''
arXiv:hep-th/0110200.
%%CITATION = HEP-TH 0110200;%%
}

\lref\RussoTJ{
J.~G.~Russo and A.~A.~Tseytlin,
``Exactly solvable string models of curved space-time backgrounds,''
Nucl.\ Phys.\ B {\bf 449}, 91 (1995)
[arXiv:hep-th/9502038].
%%CITATION = HEP-TH 9502038;%%
}

\lref\EmparanTY{
R.~Emparan,
``Composite black holes in external fields,''
Nucl.\ Phys.\ B {\bf 490}, 365 (1997)
[arXiv:hep-th/9610170].
%%CITATION = HEP-TH 9610170;%%
}

\lref\EmparanAU{
R.~Emparan,
``Black diholes,''
Phys.\ Rev.\ D {\bf 61}, 104009 (2000)
[arXiv:hep-th/9906160].
%%CITATION = HEP-TH 9906160;%%
}

\lref\ChattaraputiEB{
A.~Chattaraputi, R.~Emparan and A.~Taormina,
``Composite diholes and intersecting brane-antibrane configurations in  
string/M-theory,''
Nucl.\ Phys.\ B {\bf 573}, 291 (2000)
[arXiv:hep-th/9911007].
%%CITATION = HEP-TH 9911007;%%
}

\lref\LiangSP{
Y.~C.~Liang and E.~Teo,
``Black diholes with unbalanced magnetic charges,''
Phys.\ Rev.\ D {\bf 64}, 024019 (2001)
[arXiv:hep-th/0101221].
%%CITATION = HEP-TH 0101221;%%
}

\lref\EmparanBB{
R.~Emparan and E.~Teo,
``Macroscopic and microscopic description of black diholes,''
Nucl.\ Phys.\ B {\bf 610}, 190 (2001)
[arXiv:hep-th/0104206].
%%CITATION = HEP-TH 0104206;%%
}

\lref\Bonnor{
W.~B.~Bonnor, ``An exact solution of the Einstein-Maxwell equations
referring to a magnetic dipole'',
Z.\ Phys.\ {\bf 190}, 444 (1966).
}

\lref\DavidsonDF{
A.~Davidson and E.~Gedalin,
``Finite Magnetic Flux Tube As A Black And White Dihole,''
Phys.\ Lett.\ B {\bf 339}, 304 (1994)
[arXiv:gr-qc/9408006].
%%CITATION = GR-QC 9408006;%%
}

\lref\DowkerUP{
F.~Dowker, J.~P.~Gauntlett, S.~B.~Giddings and G.~T.~Horowitz,
``On pair creation of extremal black holes and Kaluza-Klein monopoles,''
Phys.\ Rev.\ D {\bf 50}, 2662 (1994)
[arXiv:hep-th/9312172].
%%CITATION = HEP-TH 9312172;%%
}

\lref\DowkerBT{
F.~Dowker, J.~P.~Gauntlett, D.~A.~Kastor and J.~Traschen,
``Pair creation of dilaton black holes,''
Phys.\ Rev.\ D {\bf 49}, 2909 (1994)
[arXiv:hep-th/9309075].
%%CITATION = HEP-TH 9309075;%%
}

\lref\DowkerGB{
F.~Dowker, J.~P.~Gauntlett, G.~W.~Gibbons and G.~T.~Horowitz,
``The Decay of magnetic fields in Kaluza-Klein theory,''
Phys.\ Rev.\ D {\bf 52}, 6929 (1995)
[arXiv:hep-th/9507143].
%%CITATION = HEP-TH 9507143;%%
}

\lref\DowkerSG{
F.~Dowker, J.~P.~Gauntlett, G.~W.~Gibbons and G.~T.~Horowitz,
``Nucleation of $P$-Branes and Fundamental Strings,''
Phys.\ Rev.\ D {\bf 53}, 7115 (1996)
[arXiv:hep-th/9512154].
%%CITATION = HEP-TH 9512154;%%
}

\lref\HawkingII{
S.~W.~Hawking, G.~T.~Horowitz and S.~F.~Ross,
``Entropy, Area, and black hole pairs,''
Phys.\ Rev.\ D {\bf 51}, 4302 (1995)
[arXiv:gr-qc/9409013].
%%CITATION = GR-QC 9409013;%%
}

\lref\CostaNW{
M.~S.~Costa and M.~Gutperle,
``The Kaluza-Klein Melvin solution in M-theory,''
JHEP {\bf 0103}, 027 (2001)
[arXiv:hep-th/0012072].
%%CITATION = HEP-TH 0012072;%%
}

\lref\BergmanKM{
O.~Bergman and M.~R.~Gaberdiel,
``Dualities of type 0 strings,''
JHEP {\bf 9907}, 022 (1999)
[arXiv:hep-th/9906055].
%%CITATION = HEP-TH 9906055;%%
}

\lref\MyersUN{
R.~C.~Myers and M.~J.~Perry,
``Black Holes In Higher Dimensional Space-Times,''
Annals Phys.\  {\bf 172}, 304 (1986).
%%CITATION = APNYA,172,304;%%
}

\baselineskip 18pt plus 2pt minus 2pt

\Title{\vbox{\baselineskip12pt 
\hbox{hep-th/0111177}
\hbox{HUTP-01/A058} 
%\hbox{CERN-TH/2001-???}
 }}
{\vbox{\centerline{From $p$-branes to fluxbranes and back}
}}
\centerline{Roberto Emparan$^a$\footnote{$^*$}
{{\tt roberto.emparan@cern.ch}. Also at Departamento de F{\'\i}sica
Te\'orica, Universidad del Pa{\'\i}s Vasco, Bilbao, Spain} and Michael
Gutperle$^b$\footnote{$^\dagger$}{\tt gutperle@riemann.harvard.edu}}
\medskip\centerline{\it $^a$Theory Division, CERN, CH-1211 Geneva 23,
Switzerland} \centerline{\it $^b$ Jefferson Physical Laboratory,  
Harvard University, Cambridge,
MA 02138, USA}

\vskip .3in \centerline{\bf Abstract}
In this note we study aspects of the interplay between fluxbranes and
$p$-branes. We describe how a fluxbrane can be physically realized as a
limit of a brane-antibrane configuration, in a manner similar to the way
a uniform electric field appears in between the plates of a capacitor.
We also study the evolution
 of a fluxbrane after nucleation of p-branes. We find that Kaluza-Klein
 fluxbranes do relax by forming brane-antibrane pairs or spherical branes,
 but we also find that for fluxtubes with dilaton coupling in a different
 range, the field strength does not relax, instead it becomes stronger
 after each nucleation bounce.  We speculate on a possible runaway
 instability of such fluxtubes an an eventual breakdown of their classical
 description.

\noblackbox

\Date{November 2001}

\newsec{Introduction}

In recent times two classes of localized solutions of General Relativity
and of the supergravity theories that derive from string/M-theory have
become the subject of a detailed study. The most prominent one consists of
black holes (in a broad sense which includes their possibly singular
charged extremal limits) and their higher dimensional counterparts,
$p$-branes. Another class consists of self-gravitating bundles of lines
of flux, generically known as fluxbranes. The oldest known fluxbrane is
a solution in Einstein-Maxwell theory known as the Melvin universe
\melvin. It describes a fluxtube where a finite amount of flux is
confined by its own self gravity. This solution can be generalized
\gibbonsb\gibbonsa\ and in recent months such fluxbranes have found considerable
attention in string theory \SaffinKY\GutperleMB\RussoTF\CostaIF
\EmparanRP\SparksGC\SaffinJG\BrecherXJ\MotlDJ\SuyamaNE\UrangaDX\SuyamaGD
\TakayanagiJJ\RussoNA\ChenNR\FigueroaNX\DudasUX\TakayanagiAJ.

Since $p$-branes are typically sources for gauge fields, while
fluxbranes consist of force lines of these gauge fields, it is clear
that there must be interesting dynamics arising from their interactions.
In this note we analyze two aspects of the interplay between them. We
will exhibit how fluxbranes can arise from brane-antibrane
configurations. One may then consider $p$-branes as the primary objects
of the theory, and our result gives a physical realization of a
fluxbrane in terms of them---fluxbranes from $p$-branes. On the other
hand, a typical fluxbrane is unstable to spontaneous formation of
spherical branes or brane-antibrane pairs--- $p$-branes from fluxbranes.
This can be seen as the mechanism by which a fluxbrane relaxes. We
aim to study how, and even if, this relaxation takes place.

While it is possible to consider fluxbrane configurations involving more
than one gauge field \RussoTJ\EmparanTY, for simplicity we will only
study the case of a single $U(1)$ field. Consider the four dimensional
Einstein-Maxwell-Dilaton system with an arbitrary dilaton coupling $a$,
\eqn\acdil{S= {1\over 16 \kappa} \int d^4 x \sqrt{-g}\big( R-
2\partial_\mu\phi\partial^\mu\phi - e^{2 a \phi}F^2\big).}
For $a=0$ the scalar $\phi$
decouples and we have the original Einstein-Maxwell theory, while for
$a=\sqrt{3}$ the action can be derived from the five dimensional Einstein
action via Kaluza-Klein reduction.
The dilatonic Melvin fluxtube solution for an arbitrary $a$
takes the form \gibbonsa 
\eqn\gibmaed{\eqalign{ds^2 &=\Lambda^{2\over 1+a^2}(- dt^2+ dz^2+ d\rho^2)+
\Lambda^{-{2\over 1+a^2}}\rho^2 d\varphi^2,\cr
e^{2 a(\phi-\phi_0)}&= \Lambda^{2a^2 \over 1+a^2}, \quad  A_\varphi
=e^{-a \phi_0} {b \rho^2 \over 2
\Lambda},}}
where 
\eqn\gibmaedb{\Lambda=1+{1+a^2\over 4}b^2 \rho^2.}
The solution is parameterized by $\phi_0$, the value of the dilaton at
the center of the fluxtube, and $b$, which characterizes the strength of
the magnetic field at 
the origin.  A 
peculiar feature of the Melvin solution is that the total integrated 
\eqn\tflux{Flux=\oint_{S_\infty} A_\varphi=e^{-a\phi_0}{4\pi\over
(1+a^2)}{1\over b},}
is finite and inversely proportional to $b$. Hence, in the limit $b\to
0$, even if the fieldstrength goes to zero at the center and the
metric becomes flatter, the total flux diverges.

We will restrict ourselves to fluxbranes of codimension two. Fluxbranes
of higher codimension have been studied in \SaffinKY\GutperleMB, but
their exact analytic solution is not fully known.

\newsec{Fluxbranes from brane-antibrane pairs}

An approximately uniform electric field is physically realized as the
field in between, and away from the edges of, the plates of a large
capacitor. A uniform field extending throughout all of space then
results as the limit where the size of the plates, as well as the
distance between them, grows to infinity. Here we show that, similarly,
a fluxbrane can be obtained as a limiting case of the field in between a
brane-antibrane pair.

Brane-antibrane configurations of the sort we need are explicitly known
only for the case of branes of codimension three. In four dimensions,
these solutions describe a static pair of black holes of opposite
charges; in \EmparanAU\ they were dubbed `diholes,' see also
\ChattaraputiEB\LiangSP\EmparanBB. The two oppositely charged black
holes can be submerged in a Melvin fluxbrane background, which can be
tuned so that the attraction between the black holes is balanced.
However we will also consider the general situation where the field is
not tuned to the equilibrium value. Generically, then, conical
singularities (cosmic strings) will be present.

The solution for the dihole in a Melvin fluxtube can be
expressed in several different coordinates. In Boyer-Lindquist-like
coordinates \EmparanAU\Bonnor\DavidsonDF, the metric is
given by 
\eqn\dilholem{ds^2=\Lambda^{2\over 1+a^2}\Big(-dt^2 + 
{\Sigma^{4\over 1+a^2}\over
( \Delta+(m^2+k^2)\sin^2\theta)^{3-a^2\over 1+a^2}}\big( {dr^2\over 
\Delta}+d\theta^2\big)\Big)
+{\Delta \sin^2\theta \over \Lambda^{2\over 1+a^2}}d\varphi^2\,,}
the gauge field,
\eqn\gaugef{A_\varphi= - e^{-a\phi_0}{{2\over\sqrt{1+a^2}}m r k +{1\over
2} b\big( (
r^2-k^2)^2+\Delta
k^2\sin^2\theta\big)\over \Lambda \Sigma}\sin^2\theta\,,}
and the dilaton
\eqn\dilatonf{e^{\phi-\phi_0}= \Lambda^{{a\over 1+a^2}}\,,}
where $\Lambda$ is 
\eqn\lamdef{\Lambda = {\Delta + k^2 \sin^2\theta + 2\sqrt{1+a^2}bmr k 
\sin^2\theta
+{1+a^2\over 4} b^2 \sin^2\theta \big( (r^2-k^2)^2 +\Delta k^2
\sin^2\theta\big)\over \Sigma}\,.}
Here we have also defined
\eqn\defds{\eqalign{\Delta=r^2-2 mr -k^2,\quad \quad 
 \Sigma&=r^2-k^2\cos^2\theta}.}
The parameter $m$ can be viewed as determining the mass and charge of
each of the black holes, while $k$ is related to the distance between
them\foot{See \EmparanAU\ for a detailed analysis.}. On the other hand, $b$
is related to the presence of a magnetic field and it is easy to see
that as we take $r\to \infty$ the solution \dilholem-\dilatonf\
asymptotes to the dilatonic Melvin solution \gibmaed\ with magnetic
field parameter $b$ and dilaton expectation value $\phi_0$.

Note that the Killing vector $\partial_\varphi$ for the axial symmetry has
vanishing norm at $\theta=0,\pi$ as well as at $r=r_+=m+\sqrt{m^2+k^2}$.
The proper interpretation of this fact is that the symmetry axis has two
parts: an `outer' part, where $\theta=0,\pi$ with $r_+\leq r<\infty$,
and an `inner' part, where $r=r_+$ and $0\leq \theta\leq\pi$. The
location where the two segments intersect, $r=r_+$, $\theta=0,\pi$,
corresponds to the location of two oppositely charged extremal dilatonic
black holes; in particular, for $a=0$ one finds oppositely
charged, extremal Reissner-Nordstrom black holes. In general such a
configuration can not be static, since the black holes attract. This
fact manifests itself in the presence of deficit
angles on the outer and inner axis
 \eqn\defang{\delta_{outer}= 2\pi -\Delta\varphi,\quad \delta_{inner}= 2\pi -
 \Big(1+{m^2\over k^2}\Big)^{2\over 1+a^2} \Big(1+ {\sqrt{1+a^2}bm r_+\over
 k}\Big)^{-{4\over 1+a^2}}  \Delta\varphi,}
where $\Delta\varphi$ is the angular periodicity $\varphi\sim
\varphi+\Delta\varphi$. These deficit angles represent `strings' and
`struts' which push apart or pull together the black holes. 

If we want to eliminate the conical deficits both at the inner and outer
axis, we can do so by choosing the periodicity $\Delta \varphi=2\pi$ and
the asymptotic magnetic field $b= k/(\sqrt{1+a^2}m r_+) ( -1+
\sqrt{1+m^2/k^2})$. This configuration represents a pair of oppositely
charged black holes held in an (unstable) equilibrium by an external
Melvin magnetic fluxtube. In other situations, when the field is not
tuned to the equilibrium value, we will choose $\Delta\varphi$ so as to
have $\delta_{inner}=0$, $\delta_{outer}>0$. 

We want to identify possible realizations of Melvin fluxtubes as
different limits of the solution above. To this effect, it is convenient
to transform the dihole metric into cylindrical Weyl coordinates, so
$r,\theta$ are replaced by $z,\rho$. Defining
\eqn\weyla{
R_\pm=\sqrt{\rho^2+(z\pm\sqrt{m^2+k^2})^2},}
the change of variables is accomplished by (see \EmparanBB\ for details)
\eqn\weylb{
r-m\equiv R={1\over 2}(R_++R_-),\quad \cos\theta={z\over
R}={R_+-R_-\over 2\sqrt{m^2+k^2}},\quad \rho=\sqrt{\Delta}\sin\theta.}

Let us first focus on the case of $b=0$, where the brane and antibrane
are kept apart by cosmic strings without any external fluxtube. In order
to have the axis $r=r_+$ free of conical singularities, we define 
\eqn\newphi{
\tilde\varphi=\left(1+{m^2\over k^2}\right)^{2\over 1+a^2}\varphi } 
and then identify $\tilde\varphi\sim\tilde\varphi+2\pi$.

We want the size of the black holes, as well as their separation, to
grow very large, in a manner that the magnetic field stretching between
them remains finite. This is a limit where both $m$ and $k$ grow to
$\infty$. To achieve this, scale
\eqn\limmkdil{
m\to\lambda^{3+a^2\over 1+a^2}m,\quad k\to\lambda^{2\over 1+a^2} k
}
and $\rho\to\lambda^{2\over 1+a^2}\rho$, $z\to\lambda^{2\over 1+a^2}z$,
$t\to\lambda^{2\over 1+a^2}t$.
Note we do not scale $\tilde\varphi$, so its periodicity remains
unaltered. Take now $\lambda\to\infty$. In this limit the conical
deficit for the strings outside the dihole becomes equal to $2\pi$ (i.e,
the geometry collapses), but since we stay in between the branes, this
singularity is pushed away to infinity. 
The metric becomes 
\eqn\limdih{\eqalign{
ds^2&=\left(1+{\rho^2\over k^2}\right)^{2\over 1+a^2}
\left[-\left({k\over 2m}\right)^{4\over 1+a^2}dt^2+ \left({2k\over
m}\right)^{4\over 1+a^2}
(d\rho^2+dz^2)\right]\cr
&+\left({2k\over
m}\right)^{4\over 1+a^2}{\rho^2\over 
\left(1+{\rho^2/ k^2}\right)^{2\over 1+a^2}}d\tilde\varphi^2\,,
}}
the magnetic potential,
\eqn\limpot{
A_{\tilde\varphi}=e^{-a\phi_0}\lambda^{2a^2\over 1+a^2}\left({k\over
m}\right)^{3-a^2\over 1+a^2}
{4m\over 1+\rho^2/k^2}\,,
}
and the dilaton,
\eqn\dillim{ \phi-\phi_0={1\over 1+a^2}\log\left(1+{\rho^2\over  k^2}\right)
+{2a\over 1+a^2}\log({2\lambda m\over k})\,.
}
Both $A_{\tilde\varphi}$ and $\phi$ diverge as $\lambda\to\infty$.
However, the divergence is a constant (independent of the
coordinates) that can be absorbed by shifting the dilaton, 
\eqn\newdil{
\phi_0\to \phi_0-{2a\over 1+a^2}\log({2\lambda m\over k})\,.
}
Then, after further rescaling
$\bar t=\left({k\over 2m}\right)^{2\over 1+a^2}t$, $\bar z=\left({2k\over
m}\right)^{2\over 1+a^2}z$, 
$\bar \rho=\left({2k\over m}\right)^{2\over 1+a^2}\rho$
one recovers the dilatonic Melvin fluxtube \gibmaed, with
\eqn\limbdil{
b={2\over \sqrt{1+a^2}}{1\over k}\left({m\over 2 k}\right)^{2\over
1+a^2}\,.
}
Hence we have found a physical realization of a fluxtube as a limit of
the field created by a brane-antibrane pair.

Now we study the limit where two black holes in an external
fluxtube are moved apart, $k\to\infty$, but $m$, and hence their size,
is kept finite. One might think that in this case one should recover the same
field $b$ as in the absence of the black holes. However, it is possible to
take the limit in such a way that the field
left over after the black holes have been removed actually differs from
the initial one. To do so,
after taking the limit $k\to \infty$, rescale $t, \rho$ and $z$ by a
factor of $(1+
\sqrt{1+a^2} bm)^{2/(1+a^2)}$ and choose the periodicity
\eqn\perphi{\Delta\varphi= 2 \pi (1+  \sqrt{1+a^2}\; bm )^{4\over 1+a^2}. }
It follows from \defang\ that there is a nonzero deficit angle on the outer
axis, corresponding to a cosmic string pulling the black holes apart.
However, in the limit $k\to \infty$ the outer segments are pushed to
infinity and only the inner axis with zero deficit angle remains.
The two black holes forming the dihole have been pulled to infinity and a
Melvin spacetime remains, with parameters
\gibmaed, with the parameters $\hat b, \hat \phi_0$ replacing $b, \phi_0$:
\eqn\parmva{\hat b= { b\over (1+ \sqrt{1+a^2}bm)^3}, \quad \quad
e^{\hat\phi_0}= e^{\phi_0} (1+ \sqrt{1+a^2} bm )^{ {a\over 1+a^2}}}
and the total integrated flux is given by
\eqn\tflux{Flux=\oint_{S_\infty} A_\phi=e^{-a\hat \phi_0}{4\pi\over
1+a^2}{1\over\hat b}.}
Note that $ (1+ \sqrt{1+a^2}bm)>1$ and hence the new coupling constant is
bigger.  The magnetic field strength at the center of the
fluxtube is smaller. 
 This shows that moving the diholes away has made the magnetic
field weaker. It is one of the curious features of the Melvin solution that
the total flux is getting larger. This is caused by the fact that the
magnetic flux, though weaker, can spread out further in a flatter spacetime
such that the total flux is increased.

\newsec{Fluxtube evolution through pair production bounces}

Generically, fluxbranes are unstable. Except for the cases where they
are supersymmetric, they can nucleate brane-antibrane pairs, or
spherical branes, in a manner analogous to Schwinger pair production.
One expects that fluxbranes relax via this process.

The production of black hole pairs in the
presence of a fluxtube can be described semi-classically by a
gravitational instanton bounce. This is obtained by analytic
continuation \gibbonsb\GarfinkleEQ\DowkerUP\DowkerBT\
 of the Ernst solution \ernst, which describes a pair of
black holes accelerating apart in an external magnetic field. The
dilatonic generalization of these solutions was found in
\DowkerUP\DowkerBT. 

The Ernst solution is known to asymptote, at large spatial distances, to
a Melvin fluxtube. This is the field on the axis outside the black
holes, and is naturally regarded as the fluxtube that nucleates the
pair. We will show below that the Ernst solution also approaches a
Melvin fluxtube at {\it future} asymptotic infinity\foot{By time
reversal invariance, the same field is present at past infinity.
However, the latter is not relevant when considering a pair creation
process.}. This fluxtube is the field left over between the black holes
when they have got infinitely apart, and can therefore be viewed as the
fluxtube that remains after pair production. In this way we can follow
the evolution of the fluxtube after successive bounces.

In the following we use the 
results and notation of \DowkerUP\DowkerBT. The dilatonic version of the
Ernst metric is
\eqn\enrstd{\eqalign{ds^2= {1\over A^2}{1\over (x-y)^2} \Big\{\Lambda^{2\over
1+a^2} 
\Big( F(x) G(y) dt^2 -{F(x)\over G(y)}dy^2 + {F(y)\over G(x)}
dx^2\Big)+ \Lambda^{-{2\over
1+a^2}}F(y) G(x) d\varphi^2 \Big\}}.}
The dilaton and gauge field are given by
\eqn\dilga{e^{2 a (\phi-\phi_0)}= \Lambda^{2 a^2\over 1+a^2} {F(y)\over
F(x)},\quad\quad A_\varphi= -{2 e^{-a\phi_0}\over (1+a^2) b \Lambda}\big( 1+
{1+a^2\over 2} b q x\big)  , }
where 
\eqn\lamdef{\Lambda= \big( 1+{1+a^2\over 2} b q x\big)^2 +{(1+a^2) b^2\over 4
A^2 (x-y)^2}G(x)F(x).}
The functions  $F(\xi)$ and $G(\xi)$ are defined by
\eqn\gedine{\eqalign{F(\xi)&= (1+ r_-A\xi)^{2 a^2\over 1+a^2}\cr
&= (r_- A)^{2
a^2 \over 1+a^2}(\xi-\xi_1)^{2a^2\over 1+a^2}\cr
G(\xi)&= (1+ \xi^2- r_+ A\xi^3)(1+ r_- A\xi)^{1-a^2\over 1+a^2}\cr
&= -(r_+A)(r_-A)^{1-a^2\over 1+a^2}(\xi-\xi_1)^{1-a^2\over
1+a^2}(\xi-\xi_2)(\xi-\xi_3)(\xi-\xi_4)  }}
and $q= \sqrt{r_+r_-/(1+a^2)}$.

The zeros of the fourth order polynomial $G(\xi)$ in \gedine\ are
denoted $\xi_1\leq \xi_2\leq \xi_3\leq \xi_4$. A fixed Minkowskian
signature is enforced by having $y<x$. The surface $y=\xi_2$ is a black
hole horizon and $y=\xi_3$ is an acceleration horizon. Note also that
$\xi_1=\xi_2$ corresponds to the limit of extremal black holes. The
variable $x$ is a polar angle and lies in $\xi_3\leq x\leq \xi_4$. The
absence of conical singularities at the poles $x=\xi_3,\xi_4$ requires
firstly 
\eqn\lgcond{G'(\xi_3)\Lambda (\xi_4)^{2\over 1+a^2}=
-G'(\xi_4)\Lambda(\xi_3)^{2\over 1+a^2},}
and then fixing the period of $\varphi$ to
\eqn\perph{\Delta\varphi= {4\pi (\Lambda(\xi_3))^{2\over 1+a^2}\over
G'(\xi_3)}.} 
The four parameters of the solution \enrstd-\lamdef, $A,b, r_+$ and
$r_-$, correspond, roughly\foot{The correspondence is accurate only in the
limit of small black holes, $r_+,r_-\ll 1/A,1/b$.}, to the acceleration,
magnetic fluxtube field, outer and inner horizons of the black hole,
respectively. Imposition of \lgcond\ leaves only three of
these as independent, by essentially enforcing Newton's law $mA\approx
qb$ (with $m=(r_++r_-)/2$). If the solution is continued to Euclidean
time, so as to provide an instanton for pair production, a further
restriction need be imposed, namely, the equality of the temperatures of
the black hole and acceleration horizons. This requires the black holes
to be extremal or
nearly extremal, $m\approx q$.

As explained in \DowkerUP, the Ernst
solution becomes a Melvin background for large spacelike distances. This
limit corresponds to $x\to \xi_3, y\to\xi_3$. The resulting Melvin
spacetime has parameters
\eqn\spaern{ \tilde b = { b G'(\xi_3)\over 2
\big(\Lambda(\xi_3)\big)^{3+a^2\over 
2(1+a^2)}}, \quad \quad e^{\tilde \phi_0}=  e^{ \phi_0}
\Lambda(\xi_3)^{{a\over 1+a^2}}. }
This is the field present {\it before} the black holes are nucleated.

We are interested in the `leftover' spacetime after the black holes have
accelerated away to infinity. To do so, we shall go to future asymptotic
infinity while remaining between the two black holes. The coordinates in
\enrstd\ only cover the part of the spacetime containing one black hole,
up to $y=\xi_3$. The latter is an acceleration horizon, and the
coordinate patch can be continued beyond it. For $\xi_3<y< \xi_4$, the
coordinate $y$ is timelike and $t$ is spacelike. We can access all this
region, while keeping $y<x$, by staying at the portion of the axis that
runs between the black holes, $x=\xi_4$. In this region (the `upper
wedge'), the geometry is not static: it evolves from the point of
closest approach between the black holes, at $y=\xi_3$, to the limit
where they are infinitely far apart, at $y=\xi_4$.

To obtain the form of the solution at asymptotically late times, 
we change variables as
\eqn\changvar{x= \xi_4- {\rho^2\over T^4},\quad y= \xi_4 -{1\over T^2},
\quad  t= 
{z\over T}.}
Taking the limit $T\to \infty$, one gets
\eqn\resa{\eqalign{\Lambda(x,y)&= \Lambda(\xi_4)- {(1+a^2)b^2
F(\xi_4)G'(\xi_4)\over 
4 A^2}\rho^2+o(1/T)\cr
A_\varphi&= - {2 e^{a\phi_0}\over (1+a^2)b \Lambda } \Lambda(\xi_4)^{1\over
2}+o(1/T)\cr
ds^2&= {F(\xi_4)\over A^2}\Big\{ \Lambda^{2\over 1+a^2}\Big(- G'(\xi_4) dz^2
+ {4\over G'(\xi_4)}dT^2- {4\over G'(\xi_4)}d\rho^2\Big) -
\Lambda^{-{2\over 1+a^2}}G'(\xi_4) \rho^2 d\varphi^2\Big\}\cr
&\quad  +o(1/T).}}
After a further rescaling  
\eqn\rescb{\hat \rho={2 \Lambda(\xi_4)^{1\over
1+a^2}\over A} \sqrt{F(\xi_4)\over  |G'(\xi_4)|}\rho, \quad \hat T={2
\Lambda(\xi_4)^{1\over 
1+a^2}\over A} \sqrt{F(\xi_4)\over  |G'(\xi_4)|}T, \quad  \hat z=  {
\Lambda(\xi_4)^{1\over 
1+a^2}\over A}\sqrt{F(\xi_4)  |G'(\xi_4)|},}
the $T\to \infty$ limit is a static Melvin spacetime. We denote the
new parameters with hats,
\eqn\spaernb{\hat b = { b |G'(\xi_4|)\over 2
\big(\Lambda(\xi_4)\big)^{3+a^2\over 
2(1+a^2)}}, \quad \quad e^{\hat \phi_0}=  e^{ \phi_0}
\big(\Lambda(\xi_4)\big)^{{a\over 1+a^2}}. }
We can now compare the Melvin background for large spatial distances
\spaern\ with the Melvin background after the black holes have accelerated
away \spaernb. Using \lgcond\ one finds
\eqn\bfrel{{\hat b^2\over \tilde b^2}= \left({\Lambda(\xi_3)\over
\Lambda(\xi_4)}\right)^{a^2-1\over a^2+1},\quad \quad
{e^{\hat\phi_0}\over e^{\tilde\phi_0}}=  \left({\Lambda(\xi_4)\over
\Lambda(\xi_3)}\right)^{a\over {1+a^2}}.}
Since $\xi_4> \xi_3$ it follows from \lamdef\ that
$\Lambda(\xi_4)> \Lambda(\xi_3)$, so the `leftover' coupling is always
larger than the asymptotic coupling. 
This also implies that, for $a>1$ the `leftover'
field strength decreases, $\hat b<\tilde b$. Hence, these fluxbranes do
relax after a bounce.

However, for $a<1$ the leftover fieldstrength {\it increases}, $\hat b >
\tilde b$. Contrary to expectation, the strength of these fluxtubes
appears to build up after pair creation. The fluxtube concentrates more
with each bounce. This has a striking consequence. The rate for pair
creation is approximately given, for
small black holes (which are the ones more likely to be produced) by
\eqn\rate{
\Gamma\sim e^{-{\pi m^2\over q \tilde b}}\,,
}
hence it is larger for larger initial field $\tilde b$. Since $\tilde b$
is enhanced after each bounce, pair production becomes increasingly
likely and the fluxtube starts a runaway process, by creating black hole
pairs at an ever increasing rate. 

Notice that this instability is not incompatible with the instanton
action being positive, hence yielding a suppressed rate for each bounce.
The latter implies that the build-up of the fluxtube strength will
proceed slowly. It is not incompatible either with energy conservation,
since the total energy is conserved in the pair production process
\HawkingII. The total energy of a fluxtube that extends to infinity is
infinite, so it will keep producing black hole pairs, and increasing its
field strength, until it reaches Planck-size values and its
semi-classical description ceases to be valid. For a physical fluxtube of
finite extent, one should take into account the fact that its energy
must decrease with each pair that is produced. In a sense, this
instability is reminiscent of Hawking evaporation, for which the
emission rate for a neutral black hole increases as the black hole evaporates. 

It is interesting that this property depends on the value of the dilaton
coupling $a$. This is likely correlated with the fact that the nature of
the $\varphi$ coordinate depends on the value of $a$. For $a<1$ the
circle size shrinks to zero as $\rho\to \infty$ and the space closes
off, whereas for $a>1$ it grows.

\medskip

\newsec{Higher Kaluza-Klein fluxbranes}

It is an interesting fact that for Kaluza-Klein theories the Melvin
spacetime can be obtained as a compactification of flat space on a
circle with twisted identifications \DowkerGB\DowkerSG. Starting with
$d$ dimensional flat space
\eqn\flata{ds^2= -dt^2+dx_1^2+\cdots+dx_{d-4}^2+ dr^2+ r^2 d\varphi^2
 +R^2 dy^2,}
where $y$ is periodic with period $2\pi$, one reduces along the orbits of
the Killing vector $\partial_y+ q 
\partial_\varphi$, which means that a translation $y\to y+2\pi $ is
accompanied by a rotation $\varphi\to \varphi+ 2\pi q R$. It is useful to
introduce a new single valued angular variable $\tilde \varphi= \varphi- q R y$
which has standard periodicity. In these coordinates the metric becomes
\eqn\flatb{ds^2= -dt^2+dx_1^2+\cdots+dx_{d-4}^2+ dr^2+ r^2 (d\tilde
\varphi+ q Rdy)^2 
+R^2 dy^2.} 
Using the standard formulae for Kaluza-Klein reduction,
\eqn\kkred{ds_d ^2= e^{{4\over \sqrt{d-2}}\phi}(dy+ 2A_\mu dx^\mu)^2 +
e^{-{4\over 
(d-3) \sqrt{d-2}}\phi}ds_{d-1}^2,}
 rescaling can bring the metric into the following canonical form
\eqn\melde{\eqalign{ds_{d-1}^2&=(1+ \tilde q^2 r^2)^{1\over d-3} \big(
-dt^2+dx_1^2+\cdots+dx_{d-4}^2+ dr^2+ {r^2 
\over 1+\tilde q^2 r^2}d\tilde \varphi^2\big) \cr
e^{{4\over \sqrt{d-2}}\phi}&= R^2(1+ \tilde q^2 r^2),\quad \quad
A_{\tilde \varphi}= 
{\tilde q r^2\over 2R^{d-2\over d-3} (1+ \tilde q^2 r^2)},\quad \quad
\tilde q= {q\over R^{1\over d-3}}.  }}
The total flux is given by $\oint A_{\tilde \varphi}= \pi /(q R)$. Such a
solution can be called a flux $(d-4)$-brane, since it enjoys $(d-3)$
dimensional Poincare invariance. Setting $d=11$ produces a RR-flux seven
brane in type IIA string theory. For $q= 1/R$ the twist is a rotation by
$2\pi$ which acts as $(-1)^F$ on fermions,
 which means that fermions have anti-periodic
boundary conditions along the M-theory circle. Such a M-theory
background  is believed to be type
0A theory \BergmanKM. This fact lead to the conjecture \CostaNW\ 
that a Melvin flux
seven brane with $q= 1/R$ is a dual description of type 0A.

\medskip

In dimensions higher than four, there are several ways to generalize a
black hole-anti-black hole pair. One is to add flat dimensions,
resulting in planar, codimension three brane-antibrane pairs. These are
infinite in extent. But one can also have a configuration of finite size
in the form of spherical branes. Static, $S^p$-spherical Kaluza-Klein
branes in $p+4$ dimensions were constructed in \DowkerSG. They can be
held in equilibrium by `conical branes' or by Melvin fluxbranes. In the
case of a spherical $p$-brane without any external fluxbranes, it is
easy to identify a limit where the $p$-brane grows very large in size,
while the geometry and fields in its interior approach a Kaluza-Klein
Melvin fluxbrane. This is analogous to the way in which we obtained
a fluxtube from a dihole in \limdih. 

In a similar way, it is possible to construct a solution with a pair of
planar, infinite brane and antibrane accelerating away, but, unless the
planar directions are compactified, the Euclidean action of these
solutions is infinite and therefore gives a vanishing decay rate. More
appropriately, the process of pair creation in the presence of a higher
dimensional Melvin fluxbranes is replaced by the creation of spherical
KK-branes (for $d=11$ these are D6-branes). These nucleate via an
instanton of finite action and exponentially expand after their
nucleation \DowkerSG. In the following we investigate the Minkowskian
evolution after the creation of the brane and look for the `leftover'
spacetime. 

The gravitational instanton mediating the creation of KK-branes in a Melvin
background is given by the Euclidean Myers-Perry \MyersUN\ black hole
(see \DowkerSG),
\eqn\dowme{\eqalign{ds^2&= \left( 1- {m\over r^{d-5}\Sigma}\right)dx_d^2
- {2 m k \sin^2\theta 
\over r^{d-5} \Sigma}dx_d d\varphi + {\Sigma\over r^2-k^2 - m r^{5-d}}dr^2 +
\Sigma d\theta^2\cr
&+ {\sin^2\theta\over \Sigma}\left( (r^2-k^2)\Sigma - {m\over r^{d-5}}k^2
\sin^2\theta\right)d\varphi^2 + r^2 \cos^2 \theta d\Omega_{d-4}, }}
where, again, $\Sigma= r^2- k^2 \cos^2\theta$.
The Minkowskian horizon is rotated to a Euclidean `bolt', with radius
$r_h$ defined by
\eqn\horloc{r_+^2-k^2- {m\over r_+^{d-5}}=0\,.}
The absence of a conical singularity at $r=r_+$ then determines the radius $R$ 
of  the Kaluza-Klein direction $x_d$. 
The second quantity characterizing the black hole solution is
the (analytically continued) angular momentum $\Omega$. In terms of $m$
and $k$, these are
\eqn\defomr{R= {2m r_+^{6-d}\over (d-3) r_+^2-(d-5)k^2}, \quad\quad \Omega=
{k r_+^{d-5}\over m}.}
%For $k=0$ one gets the $d$-dimensional euclidean Schwarzschild black
%hole. 
Since \dowme\ is asymptotically flat one can embed the black hole in
a Melvin fluxbrane by twisting. However the identifications have to be
globally well defined and this implies that there are two possible choices
of twist angle $q$,
\eqn\twotwists{q_{0A} = \Omega,\quad\quad q_{IIA}=
\Omega-{sgn(\Omega)\over R}\,.}
The two twists differ by a rotation of $2\pi$, which changes the
boundary conditions on  
spacetime fermions from antiperiodic to periodic.  The reduced metric for
 either $q$ can be expressed using the function
\eqn\redmeta{\eqalign{\Lambda=& R^2\Big(1- {m\over r^{d-5} \Sigma}- q {2 m k 
\sin^2 \theta\over r^{d-5}\Sigma}+ q^2 {\sin^2\theta \over \Sigma}\big( 
( r^2-k^2)\Sigma - m r^{5-d} k^2 \sin^2\theta\big)\Big).}}
Then the dilaton and gauge field are
\eqn\redmetb{\eqalign{
e^{ {4\over \sqrt{d-2}}\phi}=&\Lambda, \qquad 
A_\varphi 
= {R\over 2\Lambda}{\sin^2\theta\over \Sigma}\Big( -{m k\over r^{d-5}}+
q \big(  
( r^2-k^2)\Sigma - m r^{5-d} k^2 \sin^2\theta\big)\Big).}}
We are interested in the Minkowskian evolution of the spacetime after
the nucleation of a brane. The analytic continuation of
one of the ignorable angular variables of the $d-4$ sphere results into
a boost coordinate that then serves as the timelike coordinate after
nucleation. The Lorentzian metric post-nucleation is then given by
\eqn\redmetc{\eqalign{ds^2&= \Lambda^{{1\over d-3}}\Big\{
{\Sigma\over r^2 - k^2 - m r^{5-d}}dr^2 + \Sigma d\theta^2+ r^2
\cos^2\theta (-dt^2 + \cosh^2 t d\Omega_{d-5}^2) \cr 
&+{R^2\over \Lambda} \sin^2\theta\big(r^2-k^2 - m
r^{5-d}\big)d\varphi^2\Big\}. }} 
As explained in \DowkerSG\ this metric for the choice $q=q_{IIA}$ 
describes a spherical D6-brane
expanding in a Flux 7-brane background. It is natural to ask what the
'leftover' spacetime after the D6-brane has moved to infinity looks
like. The metric \redmetc\ has an acceleration horizon and only
covers the region of spacetime inside it. To continue past it, it is
useful to make some coordinate changes. 
Firstly define $z= r\cos\theta$ and $\tilde r= f(r) \sin \theta$, where
\eqn\fdeft{{1\over f}{d f\over dr}= {r\over r^2 - k^2 - m r^{5-d}}.}
Secondly, define Rindler like coordinates $X,T$ in terms of $z,t$ by
$z=\sqrt{X^2-T^2}$ and $t= {\rm arctanh}(T/X)$. The exact form of the metric
in the new coordinates is very complicated, but we are only
interested in the $T\to \infty$ limit. So we need simply analyze the
leading part of the solution, dropping terms of order $1/T$. 

Now, in order to get a static metric in terms
of the new coordinates, the old radial coordinate has to behave as
\eqn\scaler{r= r_+ +\left({\tilde r\over T}\right)^{1/c_h}= r_+ +
{1\over 4 c_h^2}\left({\hat r\over T}\right)^2,}
where we have defined $c_h= Rr_+^{d-4}/(2 \mu)$.
Then, as $T\to \infty$, \redmeta\ becomes
\eqn\redetal{\Lambda= R^2\Big( \big(1- {q\over \Omega}\big)^2+
{k^2\over r_+^3 c_h}{q^2\over R^2 \Omega^2}\hat r^2\Big)+o(1/T)\,,}
with the gauge field 
\eqn\redmetbl{A_\varphi= {R\over 2\Lambda}\Big( {q-\Omega\over
\Omega^2}+{k^2\over r_+^3 c_h}{q\over R^2 \Omega^2}\hat r^2\Big)+o(1/T)\,,
}
and the metric
\eqn\redmetcl{ds^2 = \Lambda^{{1\over d-3}}\Big\{ -{k^2\over r_+^2}
dT^2 + {k^2 \over r_+^3 c_h} d\hat r^2 + dX^2 + X^2 d\Omega_{d-5}^2 +
{1\over \Omega^2 \Lambda} {k^2\over r_+^3 c_h}\hat r^2 d\varphi^2\Big\}.}

Note that for $q=q_{0A}$ the constant term in \redetal\ vanishes, 
which means that the metric is not of Melvin form. This is sensible, since 
the bounce for the 0A twist corresponds to an expanding `bubble of
nothing' \WittenGJ.

For $q=q_{IIA}$ the metric can be brought into Melvin form.
After a further finite rescaling of coordinates $T \to \Omega^{1\over
d-3} {r_+/k}\; T$,
$\hat r \to \Omega^{1\over d-3} r_+^{3/2} c_h^{1/2}/k\; \hat r$,
the metric \redmetcl\ can again be brought into
the canonical form \melde\ with parameters
\eqn\finpa{\hat b= q \Omega^{1\over d-3}, \quad e^{\hat\phi_0}=
\Omega^{-{\sqrt{d-2}\over 2}}\,.}
These are the parameters of the asymptotic fluxtube at future infinity,
which we take as the field left over after nucleation of a spherical
brane. We can compare it to the (tilded) field before nucleation (i.e., the
asymptotic field at spatial infinity),
\eqn\compr{\left({\hat q \over \tilde q}\right)^2 = (R \Omega)^{2\over
{d-3}}, \quad\quad
{e^{\hat \phi_0}\over e^{\tilde \phi_0}}= (R\Omega)^{-{\sqrt{d-2}\over
2}}.}
Note that it follows from \defomr\ that $\Omega R<1$. Hence, after the
nucleated KK-brane has accelerated away, the field strength at the center
of the fluxbrane decreases, corresponding to the field being
discharged. On the other hand, the dilaton, and hence the
compactification radius, has increased.

As a check that the continuation of the coordinates past the
acceleration horizon employed in this section is correct, observe that
if we set $d=5$ and compare with the results of section 3 for
$a=\sqrt{3}$, then \bfrel\ agrees with \compr\ when $\Omega R=
(\Lambda(\xi_3)/\Lambda(\xi_4))^{1/2}$, which is indeed the case (see
\DowkerGB).

\newsec{Discussion} 

In this note we have discussed several ways in which a more complicated
spacetime involving $p$-branes becomes a fluxbrane by taking a certain
limit or looking at the long time limit of nonstatic spacetimes. First
we discussed how two black holes with opposite charges, in an otherwise
empty space, can give rise to a fluxtube as the field in between them.
We let the black holes grow very large and far apart, and in the limit
obtained an exact dilatonic Melvin fluxtube. Alternatively, we may have
viewed this limit as keeping the size of the black holes and their
distance fixed, and then focusing on the region near the middle point in
between the black holes, and at a small distance from the axis. This
region, then, is well approximated by a fluxtube.

Secondly, we discussed the spacetime describing accelerating black holes
in Melvin spacetimes and found the `leftover' spacetime after the
nucleated black holes or branes have accelerated away to infinity. For
dilaton coupling $a>1$ the fieldstrength decreases after pair
production, which we take as evidence that the field discharges.
However, we have found a striking phenomenon for dilaton coupling $a<1$.
For these cases, the strength of the magnetic field {\it grows} with
each pair that is produced. 

We pointed out that this signals a runaway instability, which is present
for the particular case of Einstein-Maxwell theory. One might then worry
that the magnetic fields in our Universe could be dangerously unstable
to exploding into a myriad black holes. However, for magnetic fields
well below the Planck scale the exponential suppression yields an
extremely slow rate for pair production, so the rate at which the field
builds up is presumably too small to have any observable
consequences.\foot{For example, for the magnetic fields at the surface
of a neutron star, the average time for producing a single black hole
pair is much larger than the age of the Universe.}

In this note we only considered nucleation processes `on the axis,'
which means that the black holes or branes are nucleated at the center
of the fluxbrane. Nucleation can also happen off the axis (possibly with
lower rate, since the field strength decreases as one moves away from
the center). We expect that the qualitative features of such processes
are similar to the ones discussed in this note. Note however that no
exact solution generalizing the Ernst solution in this respect is known
at present. 

Also, we have only studied fluxbranes of codimension two. It appears
reasonable to expect that there exist realizations of fluxbranes of
higher codimensions as limits of brane-antibrane configurations similar
to those studied in this paper. However, the required solutions are not
known. It would be very interesting to determine which of these
fluxbranes relax or not by spontaneous nucleation of $p$-branes.

Finally, we note that the case of the 11 dimensional KK-Melvin is
especially interesting due to the conjectured relation of a `critical'
Melvin with $q_{IIA}=\pm 1/R$ to ten dimensional 0A theory \CostaNW. In
this case the nucleation of D6 branes indeed relaxes the field and it is
suggestive that such a process could be interpreted as dual to the
perturbative tachyon condensation in type 0A.

\medskip
{\bf Acknowledgments}
\medskip
RE is grateful to the Perimeter Institute in Waterloo for hospitality
while part of this work was carried out, and would like to thank Rob
Myers for conversations. MG gratefully acknowledges the hospitality of
Stanford and UCSB and would like to thank S.~Giddings, R.~Gopakumar,
S.~Minwalla, M.~Strassler and especially G.~Horowitz for useful
conversations. We are also grateful to Andrew Chamblin for his
assistance during the initial stages of this project. RE acknowledges
partial support from UPV grant 063.310-EB187/98 and CICYT AEN99-0315.
The work of MG is supported in part by NSF grant NSF-PHY/98-02709. 
\listrefs  

\end